\documentclass[prc,floatfix,nofootinbib,showpacs,showkeys,reprint]{revtex4-2}

\pdfoutput=1

\usepackage[utf8]{inputenc}
\usepackage[T1]{fontenc}
\usepackage{amsmath}
\usepackage{amsfonts}
\usepackage{amssymb}
\usepackage{graphicx}
\usepackage{url}

\begin{document}

\title{Hadronization and Color Transparency}

\author{Kai Gallmeister and Ulrich Mosel}

% Affiliations / Addresses (Add [1] after \address if there is only one affiliation.)
\affiliation{Institut fuer Theoretische Physik, Universitaet Giessen, D-35392 Giessen, Germany}

\begin{abstract}
 In this article we review our work on the production of hadrons in the nuclear environment. We work with a string-breaking model for the initial production of hadrons and with a quantum-kinetic transport model (GiBUU) to describe the final state interactions of the newly formed (pre)hadrons. The latter are determined both by the formation times and by the time-development of the hadron-hadron cross section. We first show that only a linear time-dependence is able to describe the available hadronizatin data. We then compare with detailed data from HERMES and JLAB; very good agreement is reached in all reactions studied without any tuning of parameters. We also repeat predictions of spectra for pions and kaons at JLAB@12GeV. We finally discuss the absence of color transparency (CT) effects in the recent experiment on proton transparencies in quasi-elastic (QE) scattering events on nuclei. We propose to look instead for CT effects on protons in semi-inclusive deep inelastic scattering (SIDIS) events.
\end{abstract}
	
\maketitle

\section{Introduction}	
One of the interesting predictions of QCD for a nuclear physics phenomenon is that of Color Transparency (CT). Particles initially produced in a hard, high-$Q^2$, process on a nuclear target are predicted to be produced as point-like configurations (PLC) with very small transverse dimensions and, correspondingly, also very small cross sections with a surrounding medium. This should have observable consequences for hadrons traversing the nuclear target on their way out to the detector. The search for CT effects has experienced ups and downs. While some experimental results are taken as evidence for CT \cite{Dutta:2012ii}, others are less convincing or even negative such as the recent result on proton transparency in (e,e'p) QE events \cite{HallC:2020ijh}.

Experimental searches for CT have relied mostly on meson production on nuclear targets. For a given velocity of the produced particle the target radius then provides a time scale for hadronization. An often-cited case is that of the Fermilab experiment E791 that looked at the diffractive dissociation of an incoming 500 GeV pion beam into di-jets\cite{E791:2000kym} which was analyzed in terms of CT by Frankfurt, Miller and Strikman \cite{Frankfurt:2000jm}, using the Glauber approximation. Later experiments at JLAB looked both at pion and rho production as a function of $Q^2$. An up-to-date review of these experiments and their theoretical analyses can be found in \cite{Dutta:2012ii}.

One of the problems in identifying CT in such experiments lies in the fact that a reference cross section for a process without CT is needed. For this reference often the cross section on an individual nucleon or on deuterium is used. The ratio of particle production cross sections with and without final state interactions,  i.e.\ nuclear transparency, then is also influenced by 'trivial' nuclear physics effects, such as nuclear binding and Fermi motion. This complicates the identification of genuine CT effects.

The arguments leading to the prediction of CT are based on pQCD which was generally assumed to be valid for $Q^2 > 1 -2 $ GeV$^2$. Furthermore, in \cite{Collins:1996fb} it was argued that CT is predominantly caused by interactions with longitudinally polarized photons, whereas processes with transversely polarized photons should be suppressed by powers of $1/Q^2$ relative to interactions with longitudinal photons. The exact onset of such suppression is, however, quite uncertain. Experimentally it has been shown in Ref.\ \cite{Horn:2007ug} that at $Q^2$ up to $ \approx 4$ GeV$^2$ the transverse cross section for pion production is {\it larger} than the longitudinal one by about a factor of 2, contrary to the pQCD expectation. Even for the Cornell and HERMES data with much higher $Q^2$ up to 10 GeV$^2$ it was shown that the transverse cross section dominates \cite{Kaskulov:2010kf}. The simple arguments from pQCD thus have to be taken with caution \footnote{Indeed, in a recent publication \cite{Larionov:2020lzk} the authors now assume that the point like configurations are formed both for longitudinal and transverse photons.} .

The puzzle of the large transverse cross section at high $Q^2$ was solved in Ref.\ \cite{Kaskulov:2008xc,Kaskulov:2009gp,Kaskulov:2010kf} by adding a hard scattering amplitude to the $t$-channel amplitude; the latter alone produced most of the longitudinal strength. This hard-scattering amplitude was connected to the excitation of high-lying resonances, which make up the DIS contribution. The decay of these DIS configurations can be described by a string fragmentation model. 

In our studies of CT and cross-section evolution we have taken the position that CT is primarily connected with these hard transverse event. The decay times of the high-lying resonances essentially determine the formation times of the final-state hadrons. The purpose of the present paper is to review the results obtained in such an approach, so confront them with new data and to point out what we have learned about the time-development of newly formed hadrons. 

\section{Model}

\subsection{Quantum-kinetic Transport}
Any description of CT requires a reliable description of final state interactions of the newly formed hadron with the surround nuclear target. For this description we use a quantum-kinetic transport model, based on the Kadanoff-Baym equations \cite{Kad-Baym:1962}, for the description of the nuclear reaction. The theoretical basis and details of the actual implementation of this model, called GiBUU, is described in some detail in \cite{Buss:2011mx}; the code is freely available from \cite{gibuu}. The treatment of final state interactions within this theory goes well beyond the standard Glauber calculations since it allows for coupled channel effects and sidefeeding of the particular channel under investigation.

\subsection{Formation Times of Hadrons}
The fate of hadrons produced in a hard photonuclear reaction on a nuclear target is governed by their formation time and the interaction cross section until the hadron has been fully formed. The formation time is related to the inverse width of the high-lying resonances that make up the DIS doorway state. We describe the actual decay of such a high-lying state ($W > 2 - 3$ GeV), by means of the LUND string fragmentation model as it is implemented in the code PYTHIA \cite{Sjostrand:2006za} which is used by GiBUU. Within PYTHIA first a string is formed which is then fragmented into the final hadrons. The space-time four-dimensional points where the string breaks determine the production times of the quarks that will make up the final hadrons. In order to form color-neutral hadrons quarks from different breaking points will have to meet. This happens at a later so-called formation time. The authors of Ref.\ \cite{Gallmeister:2005ad} have managed to extract these two relevant times from the string-fragmentation process; a detailed discussion of these results can be found in that reference.

\begin{figure}
	\centering
	\includegraphics[width=0.9\linewidth]{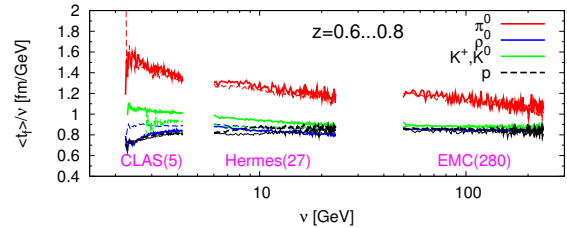}
	\caption{Average formation times in the lab frame as a function of energy transfer $\nu$ for an intermediate $z = E_h/\nu = 0.6..0.8$ in three kinematical regimes for some hadrons.}
	\label{fig:formation-times}
\end{figure}

Distributions of such times are shown in \cite{Gallmeister:2005ad}; they range from a few fermi up to well above 10 fm, depending on the initial energy transfer. The formation times also depend on mass of the produced hadron, as can be seen from Fig.\ \ref{fig:formation-times} which shows the formation times for the three kinematical regimes JLAB, HERMES and EMC for various hadrons. It is striking to see that all the heavier particles, from kaons up to protons, cluster at values of about $t_f/\nu \approx 0.8$ whereas the lighter pions are connected with larger times $t_f/\nu \approx 1.4$ at the lowest energy transfers; here $t_f$ is the formation time in the lab frame. From the results shown in Fig.\ \ref{fig:formation-times} one obtains estimates for the formation times in their restframe as a function of the mass of the produced hadron $\tau_f \approx (1 ... 2) m_h$ where $m_h$ is the mass of the produced hadron (in GeV) and $\tau_f$ is obtained in fm. 
The formation time in the lab frame is then given by $t_f = \gamma \tau_f = (0.7 .. 1.4) \nu$ ; the numerical factors (in fm/GeV) are weakly dependent on $\nu$. It is worth mentioning that, at a given energy transfer $\nu$, the formation time in the particle's restframe is the larger the larger the hadron's mass is, whereas that time in the lab frame is the smaller the larger the mass is. This is a simple consequence of a Lorentz boost when going from the rest frame to the lab frame. In particular, at a given $\nu$ the formation time for the proton is less than that of the pion as shown in Fig.\ \ref{fig:formation-times}.
 
 In our calculations we use these times as they are extracted from the PYTHIA code for every single particle and for every single event \cite{Gallmeister:2005ad}; they are not free parameters. The formed hadrons are assumed to interact with their full, normal cross section after their formation time. 

\subsection{Prehadronic Cross Sections}
For a description of CT one must specify the cross sections of the prehadrons during the production and formation time. In many high-energy event generators it is simply assumed that the initially formed PLC undergo no (or tuned attenuated) interactions until some formation time which is usually an adjustable parameter in these generators \cite{Campbell:2022qmc}.

Already in 1991 Dokshitzer et al.\ \cite{Dokshitzer:1991ab} discussed the problem how the expansion from an initially compact system to the physical hadron takes place. In their 'classical expansion model' the transverse dimension of a newly formed hadron rises linearly with time and the cross section then becomes quadratically dependent on time. Such a classical expansion model neglects the quantum-mechanical uncertainty principle which requires very large momenta for a very compact initial state of the hadron. In their 'quantum expansion model' which takes the uncertainty principle into account  Dokshitzer et al.\ arrive at a cross section that rises linearly with time. These authors finally conclude 
\begin{quote}
	"A good, complete experimental program studying almost exclusive reactions in nuclei should be able to tell us which is the better formula at a given momentum transfer."
\end{quote}

In the present calculations we explore the effects of a a small constant cross section before the formation time, as often used in generators, as well as both a linear and a quadratic time-dependence, as discussed by Dokshitzer et al. In all scenarios the prehadronic cross section is 0 before the production time and assumes the full, asymptotic value after the formation time.

The time-dependences explored thus cover  \cite{Gallmeister:2007an}
\begin{eqnarray}    
	\sigma^*(t) &=& 0.5 * \sigma_0   \label{sigma_0.5}  \\
\label{sigma_linear}
	\sigma^*(t) &=& \sigma_0\left( \frac{t-t_p}{t_f - t_p}\right)^n \quad {\rm with}\quad n = 1,2 \\
 \label{t-dep}
	\sigma^*(t) &=& \sigma_0 \left[X_0 + (1 - X_0) \cdot \frac{t - t_p}{t_f - t_p}\right] \\
                & & \text{with} \quad X_0 = n_{\rm lead} \frac{k}{Q^2}  ~.
\end{eqnarray}
Here $\sigma^*(t)$ and $\sigma_0$ are the time dependent prehadronic cross section and the final hadronic cross section, resp.
The time-dependence of Eq. (\ref{t-dep}) is very similar to that proposed by Farrar et al.\ \cite{Farrar:1988me}.
Note that only the 'pedestal' value $X_0$ explicitly depends on $1/Q^2$. The constant $k$ is chosen to be 1 GeV$^2$ and the quantity $n_{\rm lead}$ gives the ratio of the number of 'leading' quarks to total number of quarks (2 for a meson, 3 for a baryon). 'Leading' quarks are those that are connected directly with the hard interaction point; they are the endpoints of the initial string. 'Leading' hadrons are those that have at least one leading quark; they have production time $t_p =0$. For very large $Q^2$ the pedestal value $X_0$ becomes small and the time-developments of Eqs.\ (\ref{sigma_linear}) (for $n=1$) and (\ref{t-dep}) are essentially identical; this is also the case for all final hadrons that do not contain any leading quarks, i.e. that come from the inner parts of the fragmenting string. These latter particles are not connected to the hard interaction vertex and thus do not directly know about the four-momentum transfer. 

All these time-dependences of the prehadronic cross sections apply only to hard PYTHIA-generated events. The predominantly longitudinal 'QE-like' events are not affected by them.

Since $t_f \propto \nu$ at high energy transfers a large part of the hadronization happens outside the nucleus and, consequently, there is in general little sensitivity to the in-medium cross sections and to the details of the time-dependence in CT (see the discussion in \cite{Gallmeister:2007an}) whereas, in the other extreme, at low energy transfers hadronization happens very quickly and the full hadronic cross section becomes effective early on. Any observable effects of CT are thus intimately connected with the times-scales involved in the string-breaking.

\section{Results}
In any investigation of color transparency, independent of the specific reaction under study,  there are two essential properties that make CT observable:
\begin{enumerate}
\item The nuclear radius must be of the same order as (or smaller than) the distance traveled by the newly formed hadron until it reaches its final, free cross section. If that distance is much larger than the nuclear dimension most of the formation of the hadron takes place outside the nucleus and, consequently, observable effects of CT are maximized.

\item Even if the geometrical/kinematical constraint just discussed is met the actual, measurable amount of CT depends on the specific time-dependence of the cross section of the newly formed hadron with the target nucleons. 

\end{enumerate}

Both of these properties depend on the hadron's kinematics. 
Any formation time $\tau$ in the hadron's restframe is Lorentz-boosted to a larger time in the nuclear restframe. As a consequence,  for very fast hadrons, produced in high-energy collisions, the Lorentz-boost is large and most of the formation happens outside the nuclear target, again minimizing the PLC expansion and, in particular, the dependence of the attenuation on the specific time-dependence of the prehadronic cross section. It is thus essential to investigate first the nuclear modification ratio (sometimes also called 'nuclear transparency')
\begin{equation}
R_M^h(\nu, Q^2, z_h,p_T^2,\dots)= \frac{\
	\left[N_h(\nu,Q^2,z_h,p_T^2,\dots)/N_e(\nu,Q^2)\right]_A\ }{\
	\left[N_h(\nu,Q^2,z_h,p_T^2,\dots)/N_e(\nu,Q^2)\right]_D\ }\qquad 
\end{equation}
under different kinematical conditions \cite{Gallmeister:2007an}. Here all the hadronic spectra on the nucleus (``A'') as also on
deuterium (``D'') are normalized to the corresponding number of
scattered electrons. The energy transfer is denoted by $\nu$ and $z_h = E_h/\nu$ is the hadron's relative energy ($E_h$ is the energy of the produced hadron) .

\subsection{Time-dependence of prehadronic cross sections}

\begin{figure}
	\includegraphics[width=0.9\linewidth]{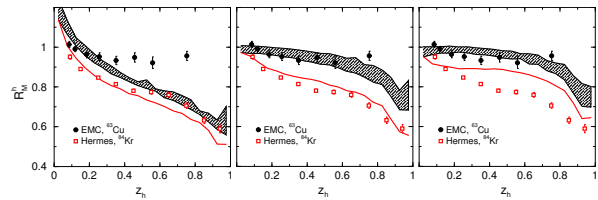}
	\caption{Nuclear modification factor for charged hadrons  as a function of relative hadron energy $z_h = E_h/\nu$, calculated by GiBUU, are compared with data from the experiments HERMES at 27 GeV beam energy \cite{Airapetian:2007vu} and EMC at 100 and 280 GeV \cite{EuropeanMuon:1991jmx}. The left figure gives results obtained with the constant prehadronic cross section, the middle figure those with a linear time-dependence and the right figure those with a quadratic dependence. The shaded band gives the theoretical prediction for 100 and 280 GeV beam energy. From \cite{Gallmeister:2007an}.}
	\label{fig:hermes-emc}
\end{figure}
The results shown in Fig.\ \ref{fig:hermes-emc} clearly demonstrate that there is a significant dependence on the special time-dependence of the prehadronic cross section; only a linear dependence can describe the data from these two very different kinematical regimes. For the EMC energies the distance traveled by the produced hadron within its formation time is considerably larger than the nuclear radius. As a consequence, the attenuation is quite small and the nuclear modification factor stays close to $R = 1$. Since in this case essentially all the hadronization takes place outside the nucleus there is very little sensitivity to the special time-dependence (linear vs quadratic). On the other hand, at energies lower than that of the HERMES experiment, e.g.\ at JLAB, the distance traveled within the formation time is small compared to the nuclear radius so that the produced hadron experiences essentially the free, asymptotic cross section.

A closer look at this data/theory comparison has shown that the pedestal $Q^2$-dependent term in Eq.\ (\ref{t-dep}) has a small, but visible influence on the modification factor \cite{Gallmeister:2007an}. In the following, we employ only the linear time-dependence with the $1/Q^2$ dependent pedestal in Eq.\ (\ref{t-dep}).

\subsection{HERMES experiment}

\begin{figure}
	\centering
	\includegraphics[width=0.9\linewidth]{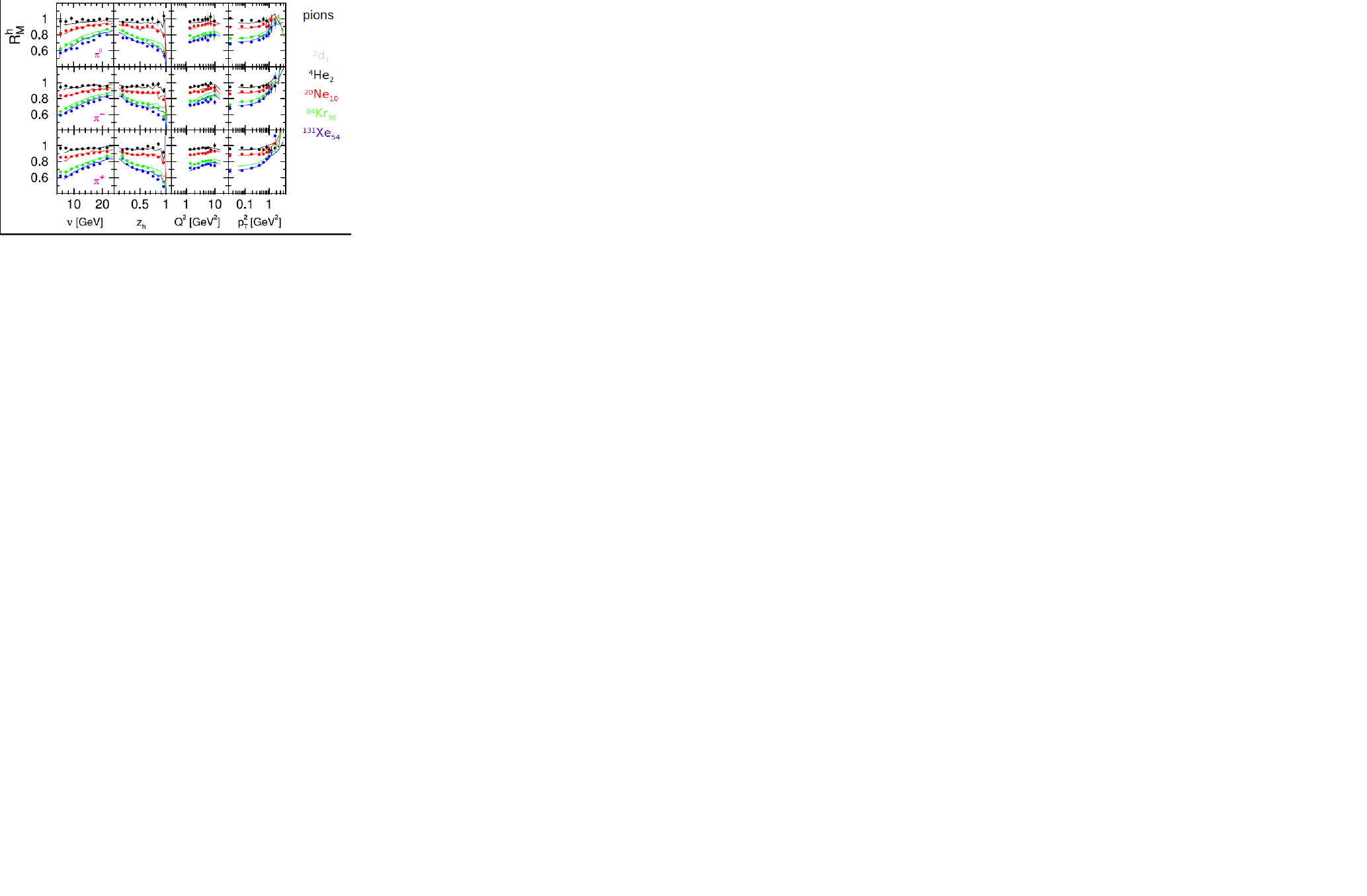}
	\caption{Nuclear modification factor for pions as a function of energy transfer $\nu$, of relative final hadron energy $z_h = E_h/\nu$, of $Q^2$ and of the the tranverse momentum $p_T^2$. The various targets are indicated in the figure. Data are from the HERMES experiment \cite{Airapetian:2007vu}.}  
	\label{fig:hermes-pions}
\end{figure}

A more detailed check of this theory for the modification factors for identified hadrons was performed in \cite{Gallmeister:2007an}.  As an illustration for the results obtained there we show in Fig.\ \ref{fig:hermes-pions} the modification factors for pions produced in the HERMES experiment. A general feature of these distributions is that they are all $<1$. This is an effect of detector acceptance that can only be described in calculations that give complete information about the final state. The actual shape, however,is a consequence of the time-dependence of the prehadronic cross sections. The agreement in all four kinematical variables $\nu, z_h, Q^2, p_T^2$ and for different target nuclei, from light to heavy, is very good. Even the two-hadron correlations could be described very well while models based on partonic effects fail to describe the correct target-mass dependence (see Fig.\ 3 in \cite{Bianchi:2007hg}).
Essential for this comparison is that the full final state is modeled so that experimental acceptances can be taken into account. We have shown in \cite{Falter:2004uc} that these experimental acceptances have a major influence in particular on the $z_h$ and $\nu$ distributions.

\subsection{JLAB experiments}
\subsubsection{JLAB@5GeV}
In Ref.\ \cite{Gallmeister:2007an} also predictions for the nuclear attenuation of pions and kaons produced at JLAB with a 5 GeV beam were given. A noticeable feature of the distributions shown in Fig.\ 6 of \cite{Gallmeister:2007an} was that the modification factor became larger than $R = 1$ for low hadron energies. As mentioned above this is not due to any time-dependence of the prehadronic cross section and it is thus not related to CT. Instead, it is simply a conseqence of final state interactions that distribute the initial energy of the primarily produced hadron on other decay products and scattering partners, a so-called 'nuclear avalanche effect'. 

Pion production data on nuclear targets were obtained with the 5 GeV beam at JLAB \cite{Clasie:2007aa} for four-momentum transfers up to about 4 GeV$^2$. The pion transparencies obtained there rise with $Q^2$ and this rise could be very well reproduced by the GiBUU calculations \cite{Kaskulov:2009gp}. Essential input for these calculations was a model for the elementary pion production cross section that allowed for a separation of longitudinal and transverse cross sections \cite{Kaskulov:2008ej} on the nucleon. Following the philosophy outlined above CT was included only for the transverse contribution. An essential feature of this calculation is thus that both the cross sections on the nucleus and those on the nucleon, which are both needed for the nuclear modification factor, are consistently calculated. Other theoretical descriptions \cite{Larson:2006ge,Cosyn:2007er} did not model the elementary cross section, but used experimental values in a Glauber calculation. In effect this means that these authors applied CT both to the longitudinal and the transverse amplitudes. Also, in these works the formation times entering into the prehadronic cross section were educated guesses only.

Similar to these pion production experiments was an experiment at JLAB looking for CT in electroproduction of $\rho$ mesons \cite{CLAS:2012tlh}. The results of this experiment also show a nuclear modification factor that increases with $Q^2$ (up to only about 2 GeV$^2$); in simple Glauber calculations this behavior has  been explained by assuming CT in both the transverse and longitudinal constributions \cite{Frankfurt:2008pz,Cosyn:2013qe}.
The data could also be explained by a calculation in which CT is only active in the transverse channel \cite{Gallmeister:2010wn}. The latter calculation again analyzed first the elementary cross section and found a significant hard component on to top of the diffractive production. CT was then applied to the hard component only.

A problem in judging the results of this experiment as evidence for color transparency lies in the fact that the experiment applied various kinematical cuts, with the purpose to exclude the resonance region and to select exclusive rho production, among others. It was shown in Ref.\ \cite{Gallmeister:2010wn} that these cuts affect the cross section on the nucleon and on the nucleus differently, mainly because of Fermi motion. Thus, the transparency as a function of $Q^2$ rises steeply for $Q^2 > 2.5$ GeV$^2$ already as a consequence of these cuts alone, even in the absence of any CT. It then becomes a quantitative question how much of an observed effect below $Q^2 = 2$ GeV$^2$ is due to CT and how much to the cuts. This question is discussed in detail in \cite{Gallmeister:2010wn}. A clean verification of CT in $\rho$ meson production in an experiment without these kinematical cuts is still outstanding.

In Ref.\ \cite{Gallmeister:2007an} a number of predictions were made for hadronization experiments at JLAB. Fig.\ 6 in that paper showed in particular predictions for the $z_h$-dependence of kaons. This predicted behavior has indeed been observed by now; the data are in very good agreement with the prediction \cite{CLAS:2011oae}. Very recently a JLAB experiment has obtained pion production data from semi-inclusive DIS events on various nuclear targets \cite{CLAS:2021jhm}. The authors there performed a detailed comparison with GiBUU calculations and found overall very good agreement of theory and experiment. Discrepancies at the lowest $z_h$ where the theory yields higher $R$ values than experimentally observed may contain interesting information on in-medium corrections to 'normal' hadronic cross sections.

\subsubsection{JLAB@12GeV}
\begin{figure}[h]
	\centering
	\includegraphics[width=0.8\linewidth]{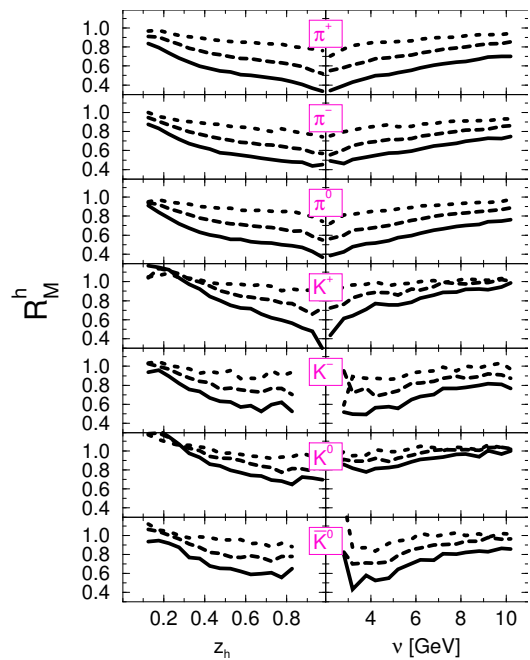}
	\caption{Predicted nuclear modification factors for pions and kaons at 12 GeV at JLAB. the short-dashed lines give results for $^{12}$C, the long-dashed line for $^{56}$Fe and the solid line for $^{208}$Pb.  From \cite{Gallmeister:2007an}.}
	\label{fig:jlab12}
\end{figure}

Since experiments with the 12 GeV beam are now running we repeat in Fig.\ \ref{fig:jlab12}  our prediction from Ref.\ \cite{Gallmeister:2007an} for pions and kaons produced on nuclear targets at that energy. The modification factors for kaons rise above $R = 1$ at small $z_h < 0.2$. This is due to the rescattering of the prehadrons. The $K^-$ attenuation is seen to be similar to that of $K^+$ because at this energy the prehadronic interactions have a strong influence. Since $K^-$ are always non-leading particles they start out with a lower prehadronic cross section; this counteracts the usual stronger absorption of $K^-$ mesons. The measurement of these kaon spectra would thus give important information on the actual production and hadronization process and the prehadronic cross sections inside the medium (see also \cite{Arleo:2003jz}).

\subsection{Proton transparency}
\begin{figure}
	\centering
	\includegraphics[width=0.9\linewidth]{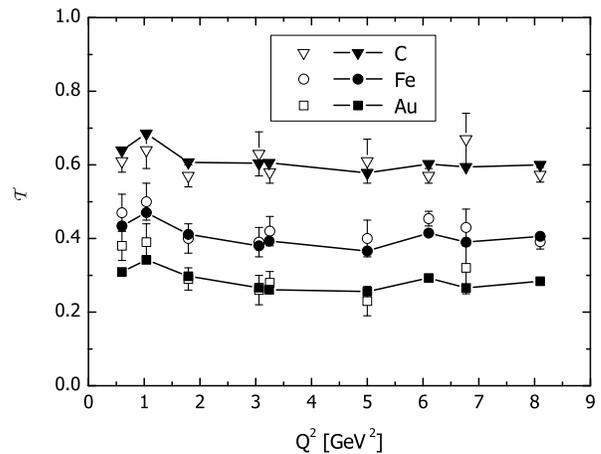}
	\caption{Transparency ratios for C, Fe and Pb targets as a function of $Q^2$. The black, solid symbols show the GiBUU results while the open symbols give the data from JLAB and SLAC (from \cite{LehrDiss:2003}, where also the references to the data can be found).}
	\label{fig:transparency}
\end{figure}

Transparency data for protons had been obtained some while ago, both at JLAB and at SLAC. An early GiBUU calculation \cite{Lehr:2001an,LehrDiss:2003} could describe these quasielastic data, which ranged up to $Q^2 = 8$ GeV$^2$, quite well (see Fig.\ \ref{fig:transparency}) without any CT effect.

Quite recently the Hall C group at JLAB published a result on the absence of CT in quasielastic $^{12}$C(e,e'p) reactions \cite{HallC:2020ijh} at even higher $Q^2$. The transparency of the protons was observed to be constant for momentum transfers $Q^2$ up to 14 GeV$^2$. This widely unexpected result led to some new theoretical studies of this problem. Miller and collaborators suggested that the results might be explained by using the so-called Feynman mechanism \cite{Caplow-Munro:2021xwi}, whereas Brodsky developed arguments why full color transparency should set in only at even higher $Q^2$  well beyond the reach of the JLAB experiment \cite{Brodsky:2021CT}. 

In the picture developed in the discussions above the explanation might be simpler: both the production and the formation time of the proton kicked out in a QE scattering event are $t_p = t_f = 0$ since a string was never stretched \footnote{This is similar to the Feynman mechanism, advocated as an explanation for the absence of CT by Miller and collaborator \cite{Caplow-Munro:2021xwi}.}. In such a situation the proton's interaction cross section with the surrounding medium does not evolve, but assumes its 'normal' value from the time of interaction, which coincides with the production time. The proton's attenuation is thus not sensitive to any time-development of the cross section and, in particular, no attenuation of the 'normal' final state interactions takes place.  This holds also for even higher $Q^2$, as long as the event is QE scattering. 

An alternative is  to look for the transparency of protons from a SIDIS event. Here, at JLAB energies, the lab formation time is about $0.8/\nu$ fm which translates into about 3 - 5 fm for $t_f$. This distance is comparable to the nuclear radius and thus effects from the linear time-dependence could be expected.  

\section{Summary and Conclusions}

In this paper we have reviewed and summarized our work on the hadronization process in deep inelastic collisions. SIDIS events on nuclear targets are sensitive to the duration (and time delays) of the hadronization process and thus give valuable information on the hadronization mechanisms.

In our work we have stressed that any such time-delays are connected only with hard, DIS-like events which require a major reorganization of the struck nucleon and the partons in the reaction product. The initial doorway state in a SIDIS event is connected with a width and thus, naturally, also with a time span for the decay.

The actual decay we handle with a string-fragmentation model as it is implemented in PYTHIA. Contrary to other approaches the production and formation times are no free parameters or educated guesses, but are directly obtained from this string-breaking process. Once these times are known a crucial property then is the time-development of the cross section experienced by the prehadrons until their formation is over. We have shown that analysis of data in very different kinematical regimes, ranging from JLAB to the EMC experiment, allows one to fix that time-dependence to be linear. The question initially asked by Dokshitzer et al \cite{Dokshitzer:1991ab} has thus been answered. 

Comparisons of these calculations with data, mainly in the HERMES and JLAB regime, have shown excellent agreement for different hadron flavors and as a function of different kinematical variables of the final state particles. We have also repeated here in this paper our old (from 2007) predictions for pions and kaons at the JLAB 12 GeV beam. Such data to compare with will appear soon.

Finally, we offer a solution to the seeming puzzle why CT was not seen in the QE scattering reactions on protons in nuclei. In a QE event formation times do not appear and thus no sensitivity to any prehadronic cross section should be expected. Instead, we propose to repeat such studies for protons from SIDIS events where observable effects are expected. 

\begin{acknowledgments}
	This work was funded by BMBF
\end{acknowledgments}

\bibliography{nuclear.bib}

\end{document}